\documentclass[twocolumn,showpacs,preprintnumbers,amsmath,amssymb]{revtex4}
\usepackage[dvips]{graphicx}
\usepackage{dcolumn}

\begin{document}

\title
{Fluctuation Conductivity and Vortex State in Superconductor with Strong Paramagnetic Pair Breaking}

\author{Naratip Nunchot, Dai Nakashima, and Ryusuke Ikeda}

\affiliation{%
Department of Physics, Kyoto University, Kyoto 606-8502, Japan
}

\date{\today}

\begin{abstract}
The fluctuation conductivity of a moderately clean type II superconductor with strong Pauli paramagnetic pair-breaking (PPB) is studied by focusing on the quantum regime at low temperatures and in high magnetic fields. First, it is pointed out that, as the PPB effect becomes stronger, the quantum superconducting fluctuation is generally enhanced so that the renormalized Aslamasov-Larkin (AL) fluctuation conductivity tends to {\it vanish} upon cooling above the irreversibility line. Further, by examining other (the DOS and the Maki-Thompson (MT)) terms of the fluctuation conductivity, the field dependence of the resulting total conductivity is found to depend significantly on the type of the vortex lattice (or, glass) ordered state at low temperatures where the strong PPB plays important roles. By comparing the present results on the fluctuation conductivity with insulating and negative magnetoresistance behaviors seen upon entering a PPB-induced novel SC phase of Iron selenide (FeSe), it is argued that the vortex matter states of the superconducting order parameter in the second lowest ($n=1$) Landau level are realized in FeSe in the parallel field configuration in high fields and at low temperatures. 

\end{abstract}



\maketitle

\section{Introduction}
Conventionally, the electrical conductivity is a reasonable measure of the effects of the critical fluctuation of the superconducting (SC) order parameter \cite{LVtext}. In particular, the resistive broadening \cite{Kwok,RI89,RI92} above a vortex lattice melting field is the most remarkable transport phenomenon in a fluctuating type II superconductor under an applied magnetic field. So far, this behavior measuring the width of the vortex liquid regime \cite{FFH,Blatter} has been studied in details in the relatively clean superconductors \cite{JPSJ601051,PRL673874} in low fields and dirty materials in a broad field range including the low temperature regime \cite{JPSJ722930}. In contrast, the fluctuation effect associated with the vortex state has been rarely discussed \cite{AdachiIkeda03} in a situation where a strong paramagnetic pair-breaking (PPB) effect is remarkably seen. 

Recently, a novel high field SC (HFSC) phase has been detected at low temperatures in the Iron-based superconductor FeSe in a field parallel (${\bf H} \perp c$) \cite{Kasa20,Hardy} and perpendicular (${\bf H} \parallel c$) \cite{Kasa21} to the basal plane of the material, and these HFSC phases are believed at present to belong to the category of Fulde-Ferrell-Larkin-Ovchinnikov (FFLO) states \cite{FF,LO} based on the anomalously high value of the ratio $|\Delta|/E_{\rm F}$ in this material, where $|\Delta|$ ($E_{\rm F}$) is a typical value of the SC energy gap (Fermi energy) \cite{MatsudaHanaguriShibauchi}. Although both of the PPB strength (the so-called Maki parameter \cite{Maki66}) and the fluctuation strength (the so-called Ginzburg number \cite{LVtext}) become larger with increasing $|\Delta|/E_{\rm F}$, a large $|\Delta|/E_{\rm F}$ in most of, e.g., heavy-fermion materials known so far did not lead to an enhanced fluctuation due to a large density of states (DOS) $N(0)$ in the normal state which reduces the Ginzburg 
number. 

Among the features on the field v.s. temperature ($H$-$T$) phase diagram of FeSe, the most remarkable ones are the diminished vortex liquid regime \cite{Hardy} above the HFSC phase and the resistive behavior \cite{Kasa20} in ${\bf H} \perp c$. There, the resistivity under a current ${\bf J}$ parallel to the field ${\bf H}$ (${\bf H} \parallel {\bf J}$) shows both an insulating behavior in the temperature ($T$) dependence and negative magnetoresistance in the field ($H$) dependence \cite{Kasa20}. The origin of this strange $T$ and $H$ dependences is believed to consist in the SC fluctuation effect because they are not seen at higher temperatures and in higher fields far above the irreversibility line and the estimated $H_{c2}$-line. It will be interesting to see whether this resistive behavior is related to the HFSC phase or not. In fact, the insulating SC fluctuation behavior has also appeared in other clean 3D materials 
\cite{Sasaki} where no behavior suggestive of strong PPB is seen, while the negative magnetoresistance behavior in the quantum fluctuation regime of a superconductor showing strong PPB effects, to the best of our knowledge, has never been reported so far. 

\begin{figure}[t]
\begin{center}
{
\includegraphics[scale=0.65]{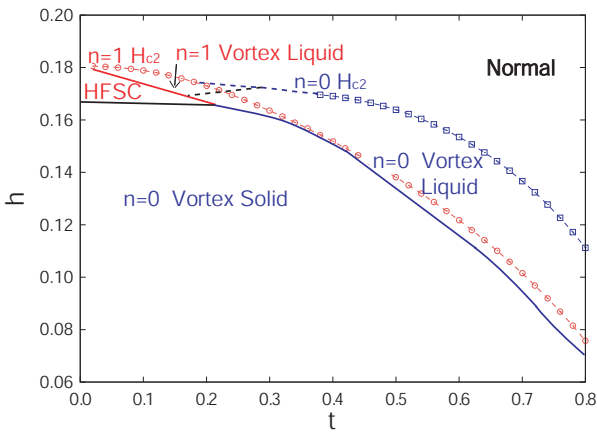}
}
\caption{(Color online) Schematic high field phase diagrams of a type II superconductor with moderately strong PPB and strong fluctuation, where $h= H/ (T_{c0}|dH_{c2}(T=T_{c0})/dT|)$, and $t=T/T_{c0}$. The blue dashed curve with square symbols denotes the mean field $H_{c2}$-curve of the SC order parameter modes in the lowest ($n=0$) LL, while the red dashed curve with circle symbols is the corresponding one of the second lowest ($n=1$) LL modes. The remaining two dashed curves are the first order $H_{c2}$-transition line (blue) and the FFLO transition line (black) for the $n=0$ LL modes in the mean field approximation. In the present situation with the SC fluctuation, all of these four dashed curves are crossover lines, and the real phase transition lines consist of the solid lines. The blue and red solid lines are the vortex glass (VG) transition lines defined in $n=0$ and $n=1$ LLs, respectively, while the black solid line expresses a structural first order transition line between the ordinary triangular pinned vortex solid (or, the vortex glass) in the $n=0$ LL and an anisotropic $n=1$ LL vortex solid \cite{Klein,Allan} which is believed to be the HFSC phase in FeSe in ${\bf H} \perp c$. We stress that all of the solid lines in this figure are just sketches of real phase transition lines. Further details are explained in 
sec.IV. 
} 
\label{Fig.1}
\end{center}
\end{figure}

In this paper, the fluctuation conductivity in an anisotropic three-dimensional type II superconductor with a moderately strong PPB effect is studied by focusing on the low $T$ and high $H$ region. First, we point out that the friction coefficient of the dissipative SC fluctuation remarkably diminishes as the PPB effect is stronger, implying an enhancement of the {\it quantum} SC fluctuation due to PPB. This qualitatively explains the insulating $T$ dependence \cite{RI96,JPSJ6533} of the resistivity curve seen in FeSe \cite{Kasa20}. 
On the other hand, the remarkable negative magnetoresistance \cite{Kasa20} in the fluctuation regime is found to be directly connected with the identity of the PPB-induced HFSC phase. One example of possible phase diagrams is shown in Fig.1, where the temperature and the magnetic field are normalized by the zero field SC transition temperature $T_{c0}$ and $T_{c0}|d H_{c2}/dT|$ at $T=T_{c0}$, respectively. It will be seen that the low $T$ portion of the $H_{c2}(T)$ curve is determined by the second lowest ($n=1$) Landau level (LL) mode of the SC order parameter. It implies that, in the case with moderately strong PPB, the SC ordered state at the low $T$ and high $H$ corner should be the vortex lattice, or the corresponding vortex glass, of the SC order parameter in the $n=1$ LL \cite{Klein,Allan,Matsu}. By comparing the fluctuation conductivity in the $n=1$ LL vortex liquid regime with that in the situation where the HFSC phase is a vortex lattice in the lowest ($n=0$) LL with a FFLO modulation parallel to the field, we find that the sum of the Maki-Thompson (MT) and DOS terms \cite{LVtext} of the fluctuation conductivity in clean limit shows negative magnetoresistance in the fluctuation regime at low enough $T$ only when the SC fluctuation is in the $n=1$ LL. 
Through the resulting qualitative agreement between the present theoretical result on the resistivity curves and the data in FeSe in ${\bf H} \perp c$, we argue that the HFSC state found in FeSe in ${\bf H} \perp c$ should be the vortex lattice or the glass formed in the $n=1$ LL. 

The present paper is organized as follows. In sec.II, the fluctuation conductivity in the Gaussian approximation is defined, and the conductivity components to be found within the Ginzburg-Landau (GL) approach are derived. In sec.III, the terms of the fluctuation conductivity to be added to the normal conductivity are investigated by focusing on the quantum regime. The resulting resistivity curves in the quantum regime are discussed based on two different scenarios in sec.IV, and the summary and our conclusion associated with the data in FeSe are mentioned in sec.V. Some details on our theoretical analysis are explained in 
Appendices.

\section{Ginzburg-Landau Fluctuation Conductivity}

We start from the weak-coupling BCS Hamiltonian 
\begin{eqnarray}
{\cal H} &=& \sum_{\sigma= \pm 1} \int d^3r \varphi^\dagger_\sigma({\bf r}) \biggl[ \frac{\hbar^2}{2m} \biggl(-{\rm i}\nabla + \frac{\pi}{\phi_0} {\bf A} \biggr)^2 - I \sigma \biggr] \varphi_\sigma({\bf r}) \nonumber \\
&-& |g| \sum_{\bf q} \Psi^\dagger({\bf q}) \Psi({\bf q}), 
\end{eqnarray}
where $\phi_0=\pi \hbar/|e|$ is the flux quantum, $I$ is the Zeeman energy, $|g|$ is the attractive interaction strength, and 
\begin{equation}
\Psi({\bf q}) = \frac{1}{2} \sum_{\bf p} \sum_{\sigma = \pm 1} \, \sigma \, w_{\bf p} \, c_{{-{\bf p} + {\bf q}/2}, -\sigma} \, c_{{\bf p}+{\bf q}/2, \sigma} 
\end{equation}
is the pair-field operator expressed by a spin-singlet pairing function $w_{\bf p}$ and $c_{{\bf p},\sigma}$ which is the Fourier transform of the electron operator $\varphi_\sigma({\bf r})$. Note that, in the mean field approximation of superconductivity, the SC order parameter field $\Delta({\bf q})$ is the ensemble average of $\Psi({\bf q})$, and, when the Cooper-pairs have a finite momentum, it is expressed by a nonvanishing ${\bf q}$. 
 
Hereafter, we assume a situation with no SC long range order. Then, $\Delta({\bf q})$ expresses the fluctuation of the SC order parameter. The total conductivity $\sigma_{zz}$ for a current parallel to the $z$-axis is given in dc limit in terms of the current-current response function $Q_{zz}$ 
by 
\begin{equation}
Q_{zz}(\Omega) = - \sigma_{zz} |\Omega|.  
\end{equation}
where 
\begin{equation}
Q_{zz}(\Omega) = \frac{1}{VT} \langle \langle j_z({\bf k}=0, \Omega) \, j_z({\bf k}=0, -\Omega) \rangle_{\rm el} \rangle_{\rm fl}. 
\label{Qzz}
\end{equation}
In eq.(\ref{Qzz}), ${\bf j}$ denotes the current density operator, the bracket $\langle \,\,\,\rangle_{\rm el}$ denotes the statistical average with respect to the electron field, while $\langle \,\,\,\rangle_{\rm fl}$ implies the ensemble average with respect to the fluctuation (boson) field $\Delta_\omega({\bf q})$, where $\Omega$ and $\omega$ are the bosonic Matsubara frequencies. 
Using the free energy $F$ for the model (1), eq.(\ref{Qzz}) can be formally written as 
\begin{equation}
Q_{zz} = - \frac{1}{V} \, \frac{\delta^2 F}{\delta A_z \delta A_z} \biggl|_{\delta A_z =0}, 
\label{QzzF}
\end{equation}
where a disturbance $\delta A_z {\hat z}$ was added to the vector potential ${\bf A}$ in eq.(1). By dividing $F$ into the normal free energy and its fluctuation contribution, the total conductivity $\sigma_{zz}$ consists of the normal conductivity $(\sigma_{\rm N})_{zz}$ to be given later (see eq.(\ref{Drude})) and the corresponding fluctuation conductivity $(\sigma_s)_{zz}$. 
As is presented in Fig.2, the three contributions (b), (d), and (e) are derived from the figure (a) expressing the Gaussian (free) fluctuation free energy based on eq.(\ref{QzzF}). 

\begin{figure}[t]
\begin{center}
{
\includegraphics[scale=1.85]{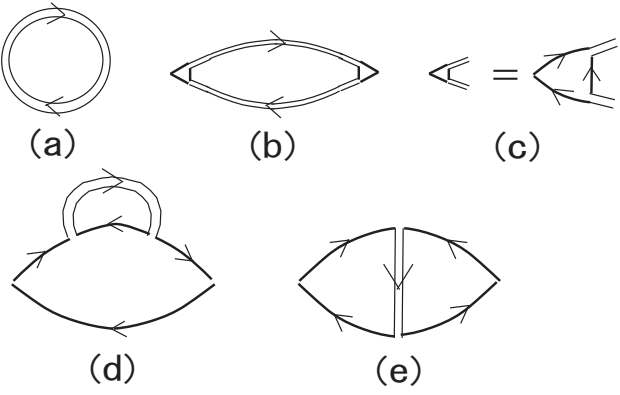}
}
\caption{(Color online) Feynman diagrams describing (a) the free energy of the Gaussian SC fluctuation, (b) $Q^{({\rm AL})}_{zz}$, (c) the triangular vertex appearing in $Q^{({\rm AL})}_{zz}$, (d) $Q^{({\rm DOS})}_{zz}|_0$, and (e) the corresponding contribution $Q^{({\rm MT})}_{zz}|_0$ from the MT diagram. A thick solid curve and a double thin solid one denote the electron Green's function and the fluctuation propagator, eq.(\ref{flucG}), respectively. 
} 
\label{Fig.2}
\end{center}
\end{figure}

In this section, we focus hereafter on the AL term drawn in Fig.2 (b) with (c) expressing the current vertex. In dc limit, the resulting conductivity $\sigma^{({\rm AL})}_{zz}$ takes the form 
\begin{equation}
\sigma^{({\rm AL})}_{zz} = - \frac{1}{|\Omega|} Q^{({\rm AL})}_{zz}(\Omega), 
\end{equation}
where 
\begin{equation}
Q^{({\rm AL})}_{zz}(\Omega) = \frac{1}{VT} \langle \langle j_z(\Omega) \rangle_{\rm el} \langle j_z(-\Omega) \rangle_{\rm el} \rangle_{\rm fl}, 
\label{QzzAL}
\end{equation}
and $j_z(\Omega)=j_z({\bf k}=0, \Omega)$. 
Equation (\ref{QzzAL}) represents the diagram Fig.2 (b). 
By using the quasiclassical treatment on the orbital contribution of the magnetic field, the averaged  current vertex, i.e., Fig.2 (c), in the expression (\ref{QzzAL}) takes the form 
\begin{eqnarray}
\langle j_z(\Omega) \rangle_{\rm el} &=& 
- \hbar v_{\rm F} \frac{\pi}{\phi_0} T \sum_{\varepsilon} \sum_{\omega} \sum_\sigma \int d\epsilon_p N(0) \nonumber \\ 
&\times& \biggl\langle |w_{\bf p}|^2 {\hat p}_z  
\int d^3r \, \Delta^*_\omega({\bf r}) {\cal G}_{\omega -\epsilon, {-\sigma}}({\bf p}) \\ 
&\times& {\cal G}_{\epsilon, {\sigma}}(-{\bf p}-{\bf \Pi})  {\cal G}_{\epsilon+\Omega, {\sigma}}(-{\bf p}-{\bf \Pi}) \biggr\rangle_{\hat p} \Delta_{\omega+\Omega}({\bf r}), \nonumber 
\label{avej}
\end{eqnarray}
where ${\bf \Pi} = - i \nabla_{\bf r} + (2 \pi/\phi_0) {\bf A}({\bf r})$, 
\begin{equation}
{\cal G}_{\epsilon, \sigma}({\bf p}+ {\bf \Pi}) = \frac{1}{ i {\varepsilon}(1+(2 \tau |\varepsilon|)^{-1}) - \epsilon_p - {\bf v}_{\bf p}\cdot{\bf \Pi} - \sigma I} 
\end{equation}
is the normal electron's Green's function, $v_{\rm F}$ is the magnitude of the Fermi velocity, ${\bf v}_{\bf p}$ and $\tau$ are the group velocity and the life time of a normal quasiparticle, respectively, and $\langle \,\,\, \rangle_{\hat p}$ denotes the angle average over the Fermi surface. 

The $\Omega$ dependence of the response function $Q^{({\rm AL})}_{zz}$ has two origins. One is the $\Omega$ dependence carried by the fluctuation field $\Delta_{\Omega}$, and the other is that carried by the normal electrons in the current vertices eq.(\ref{avej}). In contrast to the 2D case in $T \to 0$ limit \cite{GaL}, the latter contribution is nondivergent when the transition field is approached, and, in $T \to 0$ limit, leads merely to a finite value in the present 3D case which is, in clean limit, much smaller than the corresponding limiting values of the DOS and MT terms \cite{LVtext} to be discussed in sec.III. Then, we may neglect the $\omega$ and $\Omega$ dependences in the current vertices, and, as a consequence of the gauge invariance, the resulting expression of $\sigma^{({\rm AL})}_{zz}$ becomes equivalent to the fluctuation conductivity following from the corresponding Ginzburg-Landau (GL) free energy. To see this in an applied field parallel to the $z$-axis, the order parameter $\Delta$ will be decomposed into the LLs in the form 
\begin{equation}
\Delta_{n, \omega}({\bf r}) = \sum_{k, q} C_{n,k,q, \omega} u_{n,k}(x, y) e^{i qz}, 
\label{LLeigen}
\end{equation}
where $u_{n,k}$ is a normalized eigenfunction of the $n$-th LL, and $k$ is a wavenumber measuring the LL degeneracy. Using the identity $D^{-2} = \int_0^\infty d\rho \rho \exp(-D\rho)$ for an expression appearing after the $\epsilon_p$-integral in eq.(\ref{avej}) and the relation 
\begin{eqnarray}
\int \!\!\! d^3r \!\!\!\!\!\!\!&\,&\!\!\!\!\!\! \Delta_{n, \omega}^*({\bf r}) \, e^{-i \rho {\bf v}_{\bf p}\cdot{\bf \Pi}/T_{c0}} \Delta_{n, \omega+\Omega}({\bf r}) = \sum_{k,q} e^{-iv_{z, {\bf p}} \rho q/T_{c0}} \nonumber \\
&\times& \!\! (1 - \delta_{n,1} |\mu|^2 \rho^2) e^{-|\mu|^2 \rho^2/2} C^*_{n,k,q,\omega} C_{n,k,q,\omega+\Omega}
\label{appen} 
\end{eqnarray}
(see Appendix A), eq.(\ref{avej}) is found to be rewritten in the form 
\begin{eqnarray}
\langle j_z(\Omega) \rangle_{\rm el} &=& \frac{2 \pi \hbar}{\phi_0} N(0) \sum_{n,q} \biggl(\frac{\partial}{\partial q} g_n(q) \biggr) \nonumber \\ &\times& \sum_{k, \omega} C^*_{n,k,q,\omega} C_{n,k,q,\omega+\Omega}, 
\label{avej2}
\end{eqnarray}
where $\mu$ is given in terms of the velocity ${\bf v}_{\bf p} =$ ($v_x$, $v_y$, $v_z$) by 
\begin{equation}
\mu = \frac{1}{\sqrt{2} r_H} \frac{v_y + i v_x}{T_{c0}} 
\label{mu} 
\end{equation}
with $r_H=\sqrt{\phi_0/(2 \pi H)}$, and $g_n(\Pi_z)$ is the gradient operator of the resulting Ginzburg-Landau (GL) 
action 
\begin{widetext}
\begin{equation}
{\cal S}_{\rm GL} = N(0) \int d^3r \biggl( \sum_n \biggl[ T^{-1} \sum_\omega \Delta^*_{n,\omega} ( \, \eta_n |\omega| +  a_n(0) + g_n(\Pi_z) \, ) \Delta_{n,\omega} \biggr] + \frac{\beta}{2} \int_0^{T^{-1}} d\tau |\Delta(\tau)|^4 \biggr)
\label{GLaction}
\end{equation}
\end{widetext}
for the SC fluctuation $\Delta = \sum_n \Delta_n(\tau)= \sum_{n, \omega} \Delta_{n, \omega} e^{-i \omega \tau}$. 
The gradient operator is derived microscopically in the form \cite{AdachiIkeda03} 
\begin{equation}
g_n(\Pi_z)=B_n \Pi_z^2 + C_n \Pi_z^4, 
\label{grad}
\end{equation}
where 
\begin{widetext}
\begin{eqnarray}
B_n &=& \xi_0^2 \, b_n = \frac{1}{2! \, T_{c0}^2} \int_0^\infty d\rho \rho^2 f(\rho) \langle \, v_z^2 \, |w_{\bf p}|^2 e^{-|\mu|^2 \rho^2/2} (1 - \delta_{n,1} |\mu|^2 \rho^2) \rangle_{\hat p}, \nonumber \\
C_n &=& \xi_0^4 \, c_n = - \frac{1}{4! \, T_{c0}^4} \int_0^\infty d\rho \rho^4 f(\rho) \langle \, v_z^4 \, |w_{\bf p}|^2 e^{-|\mu|^2 \rho^2/2} (1 - \delta_{n,1} |\mu|^2 \rho^2) \rangle_{\hat p}, 
\label{gradcoeff}
\end{eqnarray}
\end{widetext}
\begin{equation}
f(\rho) = 2 \pi \frac{T}{T_{c0}} \frac{{\rm cos}(2 \pi I \rho/T_{c0})}{{\rm sinh}(2 \pi T \rho/T_{c0})} \, \exp\biggl(-\frac{\rho}{\tau T_{c0}} \biggr), 
\end{equation}
where $\xi_0 = v_{\rm F}/(2 \pi T_{c0})$. 
As is well known, in low fields or at high temperatures, $b_n$ is positive, while $c_n$ tends to be negative. In such cases, the $q^4$ term in the fluctuation propagator is conventionally neglected. In contrast, at lower temperatures and in higher fields so that the PPB effect becomes stronger, the coefficients $b_n$ and $c_n$ tend to change their sign (see Fig.5 in sec.IV). 

In eq.(\ref{GLaction}), the mode-coupling (last) term was phenomenologically 
introduced. Further, here and below, the units $k_{\rm B}=1$ and $\hbar=1$ are used. Then, in the Gaussian approximation, the 
$n$-th LL fluctuation propagator is given by 
\begin{eqnarray}
{\cal D}_n(q, \omega) &=& \langle |C_{n,k,q,\omega}|^2 \rangle_{\rm fl} \nonumber \\ 
&=& \frac{T}{N(0)(\eta_n |\omega| + r_n + g_n(q))} 
\label{flucG}
\end{eqnarray}
which is $k$-independent due to the LL degeneracy. 

In the Gaussian approximation, the fluctuation is unrenormalized, and the mass parameter $r_n$ in eq.(\ref{flucG}) becomes $a_n(0)$ defined in eq.(\ref{GLaction}) which vanishes as the $H_{c2}(T)$-line is approached. In fact, the $n=0$ ($n=1$) curve in Fig.1 is the line on which $a_0(0)=0$ ($a_1(0)=0$). However, $r_n$ remains nonvanishing even in lower fields once the fluctuation is renormalized by taking account of the mode-coupling terms. Further details on how to renormalize a given $n$-th LL fluctuation will be mentioned in sec.IV. 

Hereafter, our discussion will be continued by focusing on the cases with moderately strong PPB. In this case, on cooling along the $H_{c2}(T)$-curve, $c_n$ becomes positive at an intermediate temperature, while the coefficient $b_n$ of the $q^2$-term depends on the LL indices \cite{AdachiIkeda03}: The coefficient $b_0$ becomes negative at lower temperatures, and hence, $\Delta_{n=0}({\bf r})$ tends to form a spatial modulation \cite{FF,LO} parallel to the applied magnetic field ${\bf H}$. In contrast, $b_1$ remains positive even at lower temperatures, indicating that the $n=1$ LL vortex state may not be accompanied by a spatial modulation parallel to ${\bf H}$. This difference in the sign between $b_0$ and $b_1$ plays an important role in discussing the resistivity data in FeSe \cite{Kasa20} later (see 
sec.IV). 

The microscopic expression on the friction coefficient of the $n$-th LL fluctuation is given by 
\begin{equation}
\eta_n = \frac{1}{T_{c0}} \int_0^\infty d\rho \, \rho f(\rho) \langle |w_{\bf p}|^2 ( 1 - \delta_{n,1} |\mu|^2 \rho^2) e^{-|\mu|^2 \rho^2/2} \rangle_{\hat p}. 
\label{dyncoeff}
\end{equation}
To see the PPB effect on this quantity, the 2D case under ${\bf H}$ perpendicular to the 2D plane will be considered. In this case, $|\mu|^2$ is independent of the momentum ${\bf p}$ on the Fermi surface, and, in $T \to 0$ limit, $\rho f(\rho)$ approaches ${\rm cos}(2 I \rho/T_{c0})$. Then, we easily obtain
\begin{equation}
\eta_n|_{T=0} = \sqrt{\pi} \frac{r_H}{v_{\rm F}} \exp\biggl(-4 \biggl(\frac{r_H I}{v_{\rm F}} \biggr)^2 \biggr) \biggl[ 1 + \delta_{n,1}\biggl( 8 \biggl(\frac{r_H I}{v_{\rm F}} \biggr)^2 - 1 \biggr) \biggr]. 
\label{etaT0}
\end{equation}
In the absence of PPB, i.e., in $I \to 0$ limit, eq.(\ref{etaT0}) reduces to the corresponding result in 2D clean limit (see eqs.(25) and (26) in Ref.\cite{GaL}). The crucial feature in eq.(\ref{etaT0}) is the exponential reduction factor due to finite PPB. It implies an enhancement of the quantum fluctuation effect due to PPB. 

\begin{figure}[b]
\begin{center}
{
\includegraphics[scale=2.0]{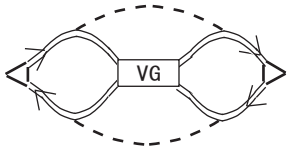}
}
\caption{(Color online) One example of the diagrams expressing $\sigma^{({\rm VG})}_{xx}$. A thick dashed line denotes a square average of the pinning potential $U({\bf r})$, a rectangle at the center denotes a VG fluctuation propagator, and the triangular vertices at the left and right edges are defined by Fig.2 (c).  
} 
\label{Fig.3}
\end{center}
\end{figure}

It is easily performed to obtain $\sigma^{({\rm AL})}_{zz}$ by substituting eq.(\ref{avej2}) into eq.(\ref{QzzAL}). Arranging the $\omega$-summation and performing the $q$-integral, the AL term of $(\sigma_s)_{zz}$ is found to become 
\begin{widetext}
\begin{eqnarray}
\sigma^{({\rm AL})}_{zz} &=& \sum_{n \geq 0} \sigma^{({\rm AL})}_{zz, \, n} = - \frac{2 \pi \xi_0}{R_Q} \frac{H}{\phi_0} \sum_{n \geq 0} \eta_n T \sum_\omega \frac{\partial}{\partial r_n(\omega)} \biggl[\frac{b_n+6\sqrt{c_n r_n(\omega)}}{2\sqrt{r_n(\omega)(b_n+2\sqrt{r_n(\omega)c_n})}} \nonumber \\
&-& \frac{2}{\sqrt{r_n}+\sqrt{r_n(\omega)}} \frac{b_n^2 - 4c_n(r_n+r_n(\omega)+\sqrt{r_nr_n(\omega)}) + 2b_n \sqrt{c_n}(\sqrt{r_n} + \sqrt{r_n(\omega)})}{(b_n-2\sqrt{c_nr_n})\sqrt{b_n+2\sqrt{r_n c_n}} + (b_n-2\sqrt{c_n r_n(\omega)})\sqrt{b_n+2\sqrt{r_n(\omega) c_n}}} \biggr], 
\label{AL}
\end{eqnarray}
\end{widetext}
where $r_n(\omega)=r_n + \eta_n|\omega|$, and $R_Q = 2 \pi \hbar/(2e)^2$ is the resistance quantum $6.45$ (k$\Omega$). The factor $H/\phi_0$ is a result of the $k$-summation, i.e., the LL degeneracy (see eq.(\ref{LLeigen})). In the case with a positive $b_n$ and a negative $c_n$, it is conventionally assumed that $c_n=0$. In this case, the corresponding GL result of the thermal fluctuation conductivity \cite{LVtext,RI89,RI92,JPSJ601051} 
\begin{equation}
\sigma_{zz}^{(cl)} = \frac{T \xi_0}{8 R_Q r_H^2} \sum_{n \geq 0} \frac{\eta_n \sqrt{b_n}}{r_n^{3/2}}
\end{equation}
is obtained as the $\omega=0$ term of eq.(\ref{AL}). On the other hand, eq.(\ref{AL}) is found, by changing the $\omega$-summation to the $\omega$-integral, to approach zero in $T \to 0$ limit and hence, leads to an increase, i.e., insulating behavior, of the resistivity \cite{RI96,JPSJ6533,JPSJ722930,Sasaki}. This quantum behavior is {\it enhanced} in the present case with strong PPB due to the above-mentioned reduction of the friction coefficient $\eta_n$. 

Next, we turn to the vortex-glass (VG) fluctuation term $\sigma^{({\rm VG})}_{zz}$ which becomes divergent as the VG transition curve $H_{\rm VG}(T)$ to be identified with the experimental irreversibility line \cite{Kasa20} is approached from above. In Ref.\cite{Hardy}, the broadened transition line of the vortex lattice melting has been determined over a wide temperature range through the heat capacity measurement in each of ${\bf H} \parallel c$ and ${\bf H} \perp c$ cases. However, it the low temperature regime of our interest where the melting line in ${\bf H} \perp c$ apparently merges with the $H_{c2}(T)$-line \cite{Hardy}, no clear data identifying the position of the melting line has been found possibly due to an effective enhancement of the vortex pinning strength at such lower temperatures. This fact implying a continuous vanishing of the resistivity there \cite{Kasa20} justifies our description of the SC transition based on the vortex-glass scaling \cite{FFH}. 

The term $\sigma^{({\rm VG})}_{zz}$ describing the vanishing of the resistivity in the vortex states is expected to diverge like 
\begin{equation}
\sigma^{({\rm VG})}_{zz} \sim \Delta_{\rm P}^3 T \biggl( \frac{H_{\rm VG}}{H-H_{\rm VG}} \biggr)^{z-1}
\label{VGscaling}
\end{equation}
with $z \simeq 4$ in 3D case \cite{FFH,DHF}. Derivation of this scaling within the $n=0$ LL fluctuation theory was performed in Ref.\cite{RI97,JPSJ722930} by focusing on ${\bf H} \perp {\bf J}$ case. It is straightforward to extend the previous analysis to the present ${\bf H} \parallel {\bf J}$ case. The outline of its derivation is described in Appendix B. 

To obtain $\sigma^{({\rm VG})}_{zz}$, an additional pinning potential term 
\begin{equation}
\delta {\cal S}_{\rm GL} = \int_0^{T^{-1}} d\tau \int d^3r \, U({\bf r}) |\Delta({\bf r}, \tau)|^2 
\label{pinning}
\end{equation}
needs to be added to the GL action (\ref{GLaction}), where the static pinning potential $U({\bf r})$ has zero mean and is assumed to obey a Gaussian ensemble, i.e., ${\overline {U({\bf r}) U({\bf s})}} = \Delta_{\rm P} \delta^{(3)}({\bf r} - {\bf s})$. Within the conventional impurity scattering model, the pinning strength $\Delta_{\rm P}$ is typically of the order of $(E_{\rm F} \tau)^{-2}$ \cite{Ishida2,GaL2}. 

As a typical term describing $\sigma^{({\rm VG})}_{zz}$ which becomes divergent according to the scaling behavior (\ref{VGscaling}), let us examine Fig.3 in which the rectangle expresses the VG fluctuation propagator to be denoted as $\chi_{\rm VG}({\bf k}; \omega_1, \omega_2)$ below (see Appendix B). By assuming only a single LL to be associated with the VG ordering and applying the formulation \cite{RI97} for describing the VG scaling (\ref{VGscaling}) to the present case, the contribution $\sigma^{({\rm VG})}_{zz}$ to the dc conductivity corresponding to Fig.3 becomes 
\begin{widetext}
\begin{eqnarray}
\sigma^{({\rm VG})}_{zz} &\simeq& - \frac{\partial}{\partial |\Omega|} \biggl(\frac{2 \pi \xi_0^2}{\phi_0} \Delta_{\rm P} \biggr)^2 \int_{\bf k} \int_{q_1} \int_{q_2} \int_{q_3}  (q_1+q_3) {\cal D}_n(q_1) {\cal D}_n(q_1+k_z) ({\cal D}_n(q_1+q_3))^2 (q_2+q_3){\cal D}_n(q_2) \nonumber \\ 
&\times& {\cal D}_n(q_2+k_z) ({\cal D}_n(q_2+q_3))^2 T \sum_\omega \chi_{\rm VG}({\bf k}; \omega, \omega+\Omega), 
\label{VGcon}
\end{eqnarray}
\end{widetext}
where inessential numerical factors are assumed to have been absorbed into the pinning strength $\Delta_{\rm P}$, and the frequency dependence was assumed to be dominated by those in the VG fluctuation propagator $\chi_{\rm VG}$. As a concrete expression of $\chi_{\rm VG}$ which is the Fourier transform of the VG correlation function, the expression derived in Ref.\cite{IshidaIkeda} 
\begin{equation}
\chi_{\rm VG}({\bf k}; \omega_1, \omega_2) = \frac{2 \, \Delta_{\rm P} \, {\overline \xi}_{\rm VG}^{2}}{2 + {\overline \xi}_{\rm VG}^2 k^2 r_H^2 + 4 {\overline \xi}_{\rm VG}^4 \eta_n(|\omega_1| + |\omega_2|) }
\end{equation}
will be used here, where ${\overline \xi}_{\rm VG}^{2} = 1/(-1 
+ H/H_{\rm VG}(T))$. 
Noting that, by arranging the $\omega$-summation in eq.(\ref{VGcon}), 
we have 
\begin{eqnarray}
T \sum_\omega [\chi_{\rm VG}({\bf k}; \omega, \omega) \!\!\! &-& \!\!\! \chi_{\rm VG}({\bf k}: \omega, \omega+\Omega)] = \biggl[ - \frac{1}{2 \pi} \chi_{\rm VG}({\bf k}; 0, 0) \nonumber \\ &+& \frac{2 {\overline \xi}_{\rm VG}^2}{\Delta_{\rm P}} \eta_n T \sum_\omega [ \chi_{\rm VG}({\bf k}; \omega, \omega) ]^2 \biggr]
|\Omega|
\end{eqnarray}
up to O($\Omega$), $\sigma^{({\rm VG})}_{zz}$ is expressed, after the analytic continuation, by 
\begin{widetext}
\begin{equation}
\sigma^{({\rm VG})}_{zz} = \frac{\Delta_{\rm P}^3 \times 10^{-3}}{R_Q \, \xi_0 \, r_n^{11/2} \, {\overline \xi}_{\rm VG}} \, \int_0^{k_c} dk k^2 \biggl[
- \frac{1}{1+k^2} + \int_{-\infty}^\infty d\varepsilon \, \varepsilon \, {\rm coth}\biggl(\frac{\varepsilon}{4 \eta_n T {\overline \xi}_{\rm VG}^4} \biggr) \frac{1+k^2}{[(1+k^2)^2 + \varepsilon^2]^2} 
\biggr], 
\label{VGcondfinal}
\end{equation}
\end{widetext}
where the cutoff $k_c$ is assumed to be a constant of order unity. In the limit of $T \gg (-1 + H/H_{\rm VG})^2/(4 \eta_n)$, this expression shows the VG scaling behavior (\ref{VGscaling}). On the other hand, in the quantum limit $T \ll (-1 + H/H_{\rm VG})^2/(4 \eta_n)$, $\sigma^{({\rm VG})}_{zz}$ vanishes like $\sigma^{({\rm AL})}_{zz}$. This vanishing should be expected. It is because Fig.3 is one of the diagrams with vertex corrections to Fig.2 (b) due to the pinning potential (\ref{pinning}), and, as shown in Ref.\cite{RI96}, any diagrams belonging to the fluctuation conductivity following from the GL action 
vanish in $T \to 0$ 
limit as far as the system is not at the VG-criticality. As mentioned above, the fluctuation conductivity terms $\sigma^{({\rm AL})}_{zz}$ and $\sigma^{({\rm VG})}_{zz}$ arising from the GL action vanish upon cooling above $H_{\rm VG}(T)$, and consequently, an insulating behavior is created in the resistivity curve. 

Before ending this section, we briefly comment on the anisotropy effects appearing in real materials which have been neglected above. Among them, one is the anisotropy in the pairing function leading to some mixing between different LL modes of the SC order parameter. For instance, in the case of a $d$-wave pairing symmetry, the $n=4$ LL modes couple to the $n=0$ LL ones. Such couplings to the higher LL modes may play a key role in resolving a subtle effect \cite{Hiasa} but do not become important in considering the global phase diagram \cite{AdachiIkeda03}. The other occurs through the Fermi velocity components reflecting the Fermi surface anisotropy. Our treatment on this anisotropy effect will be explained in sec.IV.

\begin{figure}[t]
\begin{center}
{
\includegraphics[scale=1.85]{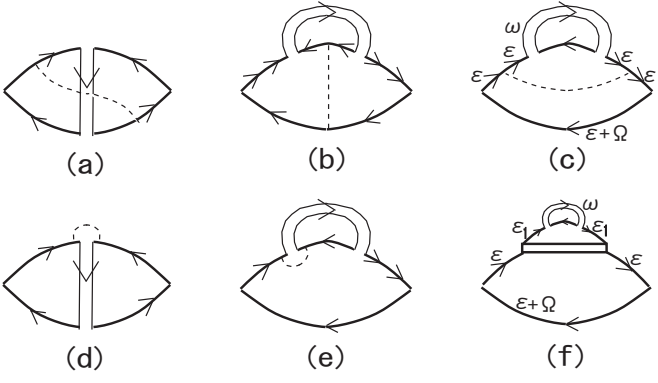}
}
\caption{(Color online) Diagrams derived from Fig.2 (d) and (e). A thin dashed line carrying the factor $1/(2 \pi N(0) \tau)$ and connecting two electron Green's functions denotes a square-averaged impurity potential, and the rectangle in (f) denotes a vertex expressing a mutual interaction between the electrons. 
} 
\label{Fig.4}
\end{center}
\end{figure}

\section{Fluctuation correction to Normal Conductivity}

Other terms of the fluctuation conductivity and related diagrams derived from them will be examined below. The DOS term, Fig.2 (d), corresponds to the self-energy correction to the familiar Drude conductivity $\sigma_{\rm N}$ 

\begin{equation}
\sigma_{\rm N} = \frac{k_{\rm F}}{3 \pi R_Q}E_{\rm F} \tau, 
\label{Drude}
\end{equation}
where $k_{\rm F}$ is the Fermi wavenumber. 
The response function $Q^{({\rm DOS})}_{zz}|_0$ corresponding to Fig.2 (d) is given by 
\begin{eqnarray}
Q^{({\rm DOS})}_{zz}|_0 &=& e^2 T \sum_\varepsilon \sum_\omega \sum_\sigma \biggl\langle \int d\epsilon_p N(0) v_z^2 |w_{\bf p}|^2 \nonumber \\
&\times& \int d^3r \langle \Delta_\omega^*({\bf r}) {\cal G}_{\varepsilon + \Omega, \sigma}({\bf p} - {\bf \Pi}) {\cal G}_{\varepsilon, \sigma}({\bf p} - {\bf \Pi}) \nonumber \\
&\times& {\cal G}_{\varepsilon, \sigma}({\bf p} - {\bf \Pi}) {\cal G}_{\omega - \varepsilon, -\sigma}(-{\bf p}) \Delta_\omega({\bf r}) \rangle_{\rm fl} \biggr\rangle_{\hat p}. 
\label{dos0} 
\end{eqnarray}
As well as in the derivation of $\sigma_{\rm N}$ \cite{AGD}, the main contribution, proportional to $\tau$, of eq.(\ref{dos0}) is obtained from the frequency region $\varepsilon(\varepsilon+\Omega) < 0$. The remaining contributions from the $\varepsilon(\varepsilon+\Omega) > 0$ region are of the same order as the neglected term in $Q^{({\rm AL})}_{zz}$ and hence, need not be considered. Then, eq.(\ref{dos0}) is straightforwardly rewritten in the form 
\begin{widetext}
\begin{eqnarray}
Q^{({\rm DOS})}_{zz}|_0 &=& 2 e^2 \frac{|\Omega|\tau T_{c0}}{1+|\Omega|\tau} \biggl\langle  \biggl(|w_{\bf p}| \frac{v_z}{T_{c0}} \biggr)^2 \int_0^\infty d\rho \, \rho \, f(\rho) \int d^3r \langle \Delta_\omega^*({\bf r}) e^{i \rho {\bf v}\cdot{\bf \Pi}/T_{c0}} \Delta_\omega({\bf r}) \rangle_{\rm fl} \biggr\rangle_{\hat p} N(0) \biggl[\frac{\tau T_{c0}}{1+|\Omega|\tau} + \rho \biggr] \nonumber \\
&=& \frac{(2e)^2 |\Omega| \tau}{2 \pi r_H^2(1+|\Omega|\tau)} 
\sum_{\omega, n \geq 0}  \int_{\bf q} N(0) {\cal D}_n(q, \omega) \biggl[ \frac{\tau T_{c0}}{1+|\Omega|\tau} \int_0^\infty d\rho \, \rho \, f(\rho) \biggl\langle |w_{\bf p}|^2 \biggl(\frac{v_z}{T_{c0}} \biggr)^2 (1 - \delta_{n,1} |\mu|^2 \rho^2) e^{-|\mu|^2 \rho^2/2} \biggr\rangle_{\hat p} \nonumber \\
&+& B_n \biggr], 
\label{Qdos0}
\end{eqnarray}
\end{widetext}
where the corresponding dc conductivity $\sigma^{({\rm DOS})}_{zz}|_0$ is given by $-2 Q^{({\rm DOS})}_{zz}|_0/|\Omega|$. However, the first term of eq.(\ref{Qdos0}) precisely cancels with one half of the corresponding MT term with $\varepsilon(\varepsilon + \Omega) < 0$ given by Fig.2 (e). Here, just as in $Q_{zz}^{({\rm DOS})}|_0$, the contribution with $\varepsilon(\varepsilon + \Omega) > 0$ was neglected in the present case where a large $\tau$ is assumed. Then, the dc conductivity resulting from the sum of Fig.2 (e) and the two diagrams corresponding to Fig.2 (d) is expressed as 
\begin{equation}
\Delta (\sigma_{\rm N})_{zz}|_0 = \frac{\tau \xi_0}{R_Q r_H^2} \sum_{n, \omega} b_n \frac{-T}{\sqrt{r_n(\omega)(b_n+2\sqrt{c_n r_n(\omega)})}}.
\label{DOSmain}
\end{equation}
The formal divergence in the $\omega$-summation will be properly cut off in performing the numerical calculation (see sec.IV). 

Other diagrams derived from Fig.2 (d) and (e) are sketched in Fig.4. In Fig.4 (a) to (e) where the life time $\tau$ is due to the elastic impurity scattering, a thin dashed line denotes an averaged impurity line. On the other hand, in the figure (f), the origin of a finite $1/\tau$ is the electron-electron scattering, and the rectangle expressing an interaction vertex takes the place of the impurity scattering. Among those additional diagrams, Fig.4 (a) and (b) vanish in the spin-singlet paired cases of our interest, because the current vertex is odd with respect to the relative momentum ${\bf p}$, while the pairing vertex $w_{\bf p}$ is even in ${\bf p}$. Below, let us discuss the roles of other diagrams in Fig.4 separately for the two cases: 

\subsection{\it {Interaction Vertex}} 

It is well known that, even in clean limit, the quasiparticle's life time $\tau$ occurs from the repulsive interaction between the quasiparticles \cite{AGD}. In general, $\tau$ is given, as the sum of the two relaxation rates due to the impurity scattering and the mutual interaction, in the form 
\begin{equation}
\frac{1}{\tau} = \frac{1}{\tau_{\rm imp}} + \frac{1}{\tau_{\rm int}}. 
\end{equation}
When the impurity scattering is negligibly weak, any vertex correction to the pairing interaction can be regarded as having been already incorporated in determining the pairing function. Then, the diagrams in Fig.2 (d), (e), and Fig.4 (f) have only to be considered as the fluctuation correction to the normal conductivity. However, Fig.4 (f) corresponding to Fig.4 (c) in the case (B) to be discussed below does not contribute to the conductivity because, as discussed in a different context \cite{Rainer}, the top section in Fig.4 (f) composed of three Green's functions with the Matsubara frequency $\varepsilon_1$ is odd with respect to $\varepsilon_1$ and hence, vanishes after taking the frequency summations. 

Therefore, as far as the quasiparticle life time is primarily determined by an interaction effect, the correction term to the normal conductivity is given by eq.(\ref{DOSmain}). 

\subsection{\it {Impurity Scattering Vertex}}

Next, we assume that the interaction effect on the quasiparticle life time is negligibly small. Then, a single dashed line in Fig.4, representing a correlator of the impurity potential, carries no finite Matsubara frequency, and the leading term of Fig.4 (c) is given by a form in which the second term of eq.(\ref{Qdos0}) is dressed. Then, eq.(\ref{DOSmain}) is replaced by 
\begin{equation}
\Delta (\sigma_{\rm N})_{zz} = \frac{\tau \xi_0}{R_Q r_H^2} \sum_{n, \omega} \delta b_n \frac{-T}{\sqrt{r_n(\omega)(b_n+2\sqrt{c_n r_n(\omega)})}} 
\label{cancelDOS}
\end{equation}
with 
\begin{eqnarray}
\delta b_n &=&  \frac{1}{2! \, T_{c0}^2} \int_0^\infty d\rho \rho^2 f(\rho) \langle ( \, v_z^2 - \langle \, (v_z')^2 \rangle_{\hat p'}) \nonumber \\
&\times&|w_{\bf p}|^2 e^{-|\mu|^2 \rho^2/2} (1 - \delta_{n,1} |\mu|^2 \rho^2) \rangle_{\hat p}, 
\label{cancelb}
\end{eqnarray}
which is the sum of Fig.2 (e) and the doubles of Fig.2 (d) and Fig.4 (c). Here, $v'_j$ ($j=x$, $y$, or $z$) denotes the $j$-component of $v_{{\bf p}'}$. 
Therefore, in the $s$-wave pairing case where $w_{\bf p}=1$, the expression (\ref{cancelb}) vanishes in 2D-like systems with ${\bf H}$ perpendicular to the plane, where $|\mu|^2$ is independent of the unit vector ${\hat {\bf p}}$, and in any system under a low enough magnetic field where the $|\mu|^2$ dependences carrying the nonlocal orbital effects of the magnetic field are negligible. 
In the $s$-wave impure case, we also need to take account of another diagrams Fig.4 (d) and (e) with the correction due to the impurity scattering to the pairing vertex. Keeping the contribution from the vertex correction of O($(\tau T_{c0})^{-1}$), it is found that the contribution of the sum of the six diagrams corresponding to Fig.4 (d) and (e) to $(\sigma_s)_{zz}$ is given by 
\begin{equation}
\Delta (\sigma_{\rm N})_{zz}|_{\rm P} = \frac{\tau \xi_0}{R_Q r_H^2} \sum_{\omega, n} \frac{T (b_n)_{\rm P}}{\sqrt{r_n(\omega)(b_n+2\sqrt{c_n r_n(\omega)})}}, 
\label{sigPPB}
\end{equation}
where 
\begin{eqnarray}
(b_n)_{\rm P}\!\!\! &=& \!\!\! \frac{1}{T_{c0}^2 \xi_0^2} \int_0^\infty d\rho \int_0^\infty d\rho' (\rho+\rho') f(\rho, \rho') \langle v_z^2 \langle (1 \nonumber \\
&-& \delta_{n,1}|\rho \mu + \rho' \mu'|^2 ) e^{-|\mu \rho + \mu' \rho'|^2/2} \rangle_{\hat p'} \rangle_{\hat p}, 
\label{ppbcoeff}
\end{eqnarray}
where $\mu'=(v_y'+iv_x')/(\sqrt{2} \, r_H T_{c0})$, and 
\begin{equation}
f(\rho,\rho') = 2 \pi \frac{T}{T_{c0}} \frac{{\rm sin}(2 I \rho/T_{c0}) {\rm sin}(2 I \rho'/T_{c0})}{{\rm sinh}(2 \pi T(\rho+\rho')/T_{c0})} e^{-\rho/(\tau T_{c0})}. 
\label{frr}
\end{equation}
Equation (\ref{frr}) implies that eq.(\ref{sigPPB}) vanishes in the absence of PPB. 

Therefore, we reach the conclusion that, in any system in zero field and a 2D-like system with no PPB and under a field perpendicular to the plane, the Gaussian fluctuation conductivity in the $s$-wave case consists only of the AL term in clean limit. This seems to be consistent with the results in clean limit in Ref.\cite{GaL,Varlamov00}. 
As is clearly seen from eq.(\ref{cancelb}), however, when the pairing function $w_{\bf p}$ has a ${\hat p}$-dependence, and/or the nonlocality $|\mu|^2$ has some momentum dependence, $\delta b_n$ remains nonvanishing with, in most cases, the same sign as $b_n$. That is, a nonvanishing contribution to the MT and DOS terms of the fluctuation conductivity in 3D case in high enough fields occurs, although its magnitude becomes generally smaller than that of eq.(\ref{DOSmain}). 
In particular, in the $s$-wave case, the different contribution $\Delta (\sigma_{\rm N})_{zz}|_{\rm P}$ induced by PPB needs to be added to $\Delta (\sigma_{\rm N})_{zz}|_0$, and, depending on the situation, the sign of the total contribution $\Delta (\sigma_{\rm N})_{zz}|_0 + \Delta (\sigma_{\rm N})_{zz}|_{\rm P}$ may change that of $\Delta (\sigma_{\rm N})_{zz}|_0$. 

In the next section, we consider two different situations belonging to the clean limit, i.e., the case in which $T_{c0}^{-1} \ll \tau_{\rm int} \ll \tau_{\rm imp}$, and the case in which $T_{c0}^{-1} \ll \tau_{\rm imp} \ll \tau_{\rm int}$.  Further, the pairing function is assumed to be dominated by a $d$-wave component \cite{Mizukami} in which the diagrams Fig.4 (d) and (e) with corrections to the pairing vertex are negligibly small. 

\section{Numerical Results} 

To compare numerical data following from the theoretical expressions in the preceding sections with experimental data, one needs to specify electronic details which are reflected in the Fermi velocity vector ${\bf v}_{\bf p}$ as well as the values of the dimensionless parameters $T_{c0}/E_{\rm F}$ and $\tau T_{c0}$. As a model on the Fermi surface, we use here a single band model and follow Ref.\cite{RI07} in which the dispersion relation of the normal quasiparticle in a quasi two-dimensional system 
\begin{equation}
\epsilon_p = \frac{1}{2m} [ p^2 + p_{\rm F}^2 {\tilde J} 
(1 - {\rm cos}(p_y d)) ] 
\label{dispersion}
\end{equation}
has been used, where $|p_y d| \leq \pi$, $d$ denotes the interlayer spacing, $p^2=p_z^2+p_x^2$, and the out-of-plane direction (i.e., $c$-axis) was chosen to be the $y$-axis. In this case, the anisotropy between the in-plane coherence length and the out-of-plane one is given by 
\begin{equation}
\gamma = \frac{2}{p_{\rm F} d} \frac{\sqrt{1 - {\tilde J}}}{\tilde J}, 
\end{equation}
and the group velocity $v_j = \partial \epsilon_p/\partial p_j$ is parametrized by $v_{\rm F}=2 \pi \xi_0 T_{c0}$, ${\tilde J}$, and $p_{\rm F} d$. 

Equation (\ref{dispersion}) is a typical model of the dispersion relation of the normal quasiparticle in a quasi two-dimensional layered superconductor. It is known that, in the case where the layer structure is important, the SC order parameter is not diagonalized in terms of the LL-basis functions in a field parallel to the layers \cite{Isotani}, and that the $n=0$ ($n=1$) LL mode weakly couples to other LLs with even (odd) $n$. As a condition justifying the neglect of the discrete layer structure and a crystalline anisotropy in describing the vortex states in the parallel field configuration, the relation 
\begin{equation}
\gamma < 1.3 \frac{\xi_0^2}{d^2} \alpha_{\rm M}, 
\label{gammaMaki}
\end{equation}
has been pointed out in Ref.\cite{RI07}. Here, $\alpha_{\rm M}$ is the Maki parameter \cite{Maki66} $\sqrt{2} H_{\rm orb}^{({\rm 2D})}(0)/H_{\rm P}(0)$, which is the measure of the strength of PPB. Then, the Zeeman energy $I$ is given by $1.62 \alpha_{\rm M} h T_{c0}$, where $h=2 \pi \xi_0^2 H/\phi_0 = \xi_0^2/r_H^2$. 
That is, if eq.(\ref{gammaMaki}) is satisfied, the discrete layer structure is unimportant so that the layered superconductor may be described by the corresponding anisotropic 3D GL model, which is obtained by replacing the gauge-invariant gradient $\Pi_j$ in eq.(\ref{GLaction}) by the following isotropized one ${\tilde \Pi}_j$, where 
\begin{eqnarray}
{\tilde \Pi}_x &=& \gamma^{1/2} {\Pi}_x, \nonumber \\
{\tilde \Pi}_y &=& \gamma^{-1/2} \Pi_y, 
\label{isop}
\end{eqnarray}
while ${\bf v}_{\bf p}\cdot{\bf \Pi}={\tilde {\bf v}_{\bf p}}\cdot{\tilde {\bf \Pi}}$ is kept. Consistently, the complex velocity $\mu$ in eqs.(\ref{appen}), (\ref{gradcoeff}), (\ref{dyncoeff}), (\ref{Qdos0}), (\ref{cancelb}), and (\ref{ppbcoeff}) needs to be replaced by 
\begin{equation}
{\tilde \mu} = \frac{{\rm i} \gamma^{-1/2} v_x + \gamma^{1/2} v_y}{\sqrt{2} T_{c0} r_H}. 
\end{equation}

Below, we focus on numerical results obtained based on the theoretical formulation in sec.II and III and 
in terms of the parameter values satisfying the relation (\ref{gammaMaki}).  
Therefore, when the temperature $T$ is normalized by the zero field SC transition temperature $T_{c0}$ defined microscopically, we only need below the dimensionless parameters $\gamma$, $\alpha_{\rm M}$, ${\tilde J}$, $T_{c0}/E_{\rm F}$, $\tau T_{c0}$, and the pinning strength $\Delta_{\rm P}$. 
In the present quasi two-dimensional case under a parallel field \cite{RI07}, the bare mass $a_n(0)$ in the GL action (\ref{GLaction}) is expressed as 
\begin{eqnarray}
a_n(0) &=& {\rm ln}\biggl(\frac{T}{T_{c0}} \biggr) + \int_0^\infty d\rho \biggl[\frac{T}{T_{c0}} \frac{2 \pi}{{\rm sinh}(2 \pi \rho T/T_{c0})}  \\
&-& f(\rho) \biggl\langle |w_{\bf p}|^2 (1 - \delta_{n,1} |{\tilde \mu}|^2 \rho^2) \exp\biggl(-\frac{|{\tilde \mu}|^2 \rho^2}{2} \biggr) \biggr\rangle_{\hat p} \biggr] \nonumber
\end{eqnarray}
for $n=0$ and $1$. Hereafter, we choose $w_{\bf p} = \sqrt{2}({\hat p}_z^2 - {\hat p}_x^2)$ by imagining that, in a model pairing function of FeSe \cite{Mizukami}, the $d$-wave pairing component dominates over 
the $s_\pm$-wave component. 

One might wonder if the present isotropized model can cover description of the case with a more complicated crystal structure in which each diagonalized order parameter mode is represented not by a single LL but by a linear combinations of the LLs. In particular, in the case with no inversion symmetry of the 
material, even a coupling between an even LL and an odd one occurs in the diagonalized modes of the SC order parameter. As shown previously \cite{Matsu,HiasaSaiki}, however, a sharp structural transition between one vortex lattice dominated by even LLs and another one dominated by odd LLs occurs even in such cases. In the familiar cases with inversion symmetry, e.g., with tetragonal crystal symmetry, a sharp transition between a state consisting only of the even LLs and dominated by the $n=0$ LL and another one consisting only of the odd LLs and dominated by the $n=1$ LL is expected to certainly occur in high fields in systems with moderately strong PPB. 

According to Ref.\cite{Kasa20}, the quantity we focus on in the following two scenarios is the field dependence of the normalized dc resistance under a current parallel to the field 
\begin{equation}
\frac{R}{R_{\rm N}} = \frac{\sigma_{\rm N}}{\sigma_{\rm N}+(\sigma_s)_{zz}}, 
\label{teikou}
\end{equation}
where $(\sigma_s)_{zz}$ is the sum of the relevant contributions to the fluctuation conductivity. The parameters $\tau T_{c0}$ and $T_{c0}/E_{\rm F}$ determine the relative magnitudes between those contributions in the fluctuation conductivity. 

To perform the fluctuation renormalization without making the procedures more complicated, the Hartree approximation \cite{RI89,RI92,Dorsey} will be used for simplicity. Then, the renormalized mass $r_n$ of the $n$-th LL fluctuation defined in eq.(\ref{flucG}) is determined selfconsistently by 
\begin{equation}
r_n=a_n(0) + 2(2n+1) \, \sqrt{2{\rm Gi}} \, h \, b_n \, \frac{T}{T_{c0}} \sum_\omega \frac{1}{\sqrt{r_n(\omega)}}. 
\label{r1}
\end{equation}
Here, the $q^4$ term in the GL action was neglected by assuming the coefficient $b_n$ of the corresponding $q^2$ term to be positive. In fact, as is shown below, the relevant fluctuation at lower temperatures is the $n=1$ LL one with the coefficient $b_1$ being positive. In eq.(\ref{r1}), the Ginzburg number 
\begin{equation}
{\rm Gi} = \frac{1}{32 \pi^2} \biggl(\frac{\beta T_{c0}}{N(0) \xi_0^3} 
\biggr)^2
\end{equation}
will be taken to be the value $6.1 \times 10^{-4}$ in the literature \cite{Koshelev}. 

For the convenience of the numerical analysis, we rewrite the self-energy term of eq.(\ref{r1}) according to 
the standard analytic continuation procedure \cite{AGD} as 
\begin{eqnarray}
\sum_\omega \frac{T}{\sqrt{r_n(\omega)}} &=& \frac{1}{\eta_n} \int_0^{\varepsilon_c} \frac{d\varepsilon}{\sqrt{2} \, \pi} {\rm coth}\biggl(\frac{\varepsilon}{2 \eta_n T} \biggr) \nonumber \\
&\times& \frac{\varepsilon}{\sqrt{(r_n^2+\varepsilon^2)(r_n + \sqrt{r_n^2 + \varepsilon^2)}}}. 
\label{hukusosekibun}
\end{eqnarray}
The cutoff $\varepsilon_c$ of the $\varepsilon$-integral will be set to a constant of order unity. The same replacement needs to be performed in eq.(\ref{DOSmain}) with $c_1=0$ if focusing on the $n=1$ LL fluctuation. 

\subsection{$T_{c0}^{-1} \ll \tau_{\rm int} \ll \tau_{\rm imp}$}

First, let us discuss numerical results obtained in the case where the impurity scattering effects are negligible so that the quasiparticle life time $\tau$ is nearly equal to $\tau_{\rm int}$. In this case, we can set 
\begin{equation}
(\sigma_s)_{zz} = \sigma_{zz}^{({\rm AL})} + \sigma_{zz}^{({\rm VG})} + \Delta (\sigma_{\rm N})_{zz}|_0.  
\label{cons0}
\end{equation}
In performing our numerical study in this case, we have chosen the parameter values ${\tilde J}=0.816$, $\gamma=1.05$, and $\tau T_{c0}=200$ in addition to the large Maki parameter $\alpha_{\rm M}=3.9$ and the large pinning strength $\Delta_{\rm P}=0.6$. 
The mean field $H_{c2}$-curves, defined as $a_n(0)=0$ \cite{Klein,Allan,Matsu,Buzdin}, for the $n=0$ and $n=1$ LL modes following from these parameter values are shown in Fig.1. Due to the large $\alpha_{\rm M}$-value, the $n=1$ LL $H_{c2}$-curve lies at higher fields in $t < 0.2$, suggesting that the $n=1$ LL fluctuation is more dominant there. Accordingly, the HFSC phase lying in lower fields is expected to be a $n=1$ LL vortex solid. Experimentally, the field-induced transition, on the black solid line in Fig.1, between the HFSC and the ordinary $n=0$ LL vortex lattice in lower fields is discontinuous consistently with our identification between the HFSC phase and the $n=1$ LL state. 

In Fig.1, the two transition lines defind within the $n=0$ LL order parameter modes and in the mean field approximation, i.e., the first order $H_{c2}$-transition and the onset of the formation of the FFLO spatial modulation parallel to ${\bf H}$, have been drawn as the blue and black dashed curves, respectively. In the real situation sketched in Fig.1 with the SC fluctuation included, both of them are crossover lines because they lie in the non SC region above the vortex-glass transition line which can be identified with the experimental irreversibility line \cite{Kasa20} or the melting line \cite{Hardy}. As pointed out previously \cite{AdachiIkeda03}, when a FFLO region expected within the mean field approximation is situated within the vortex liquid regime in the fluctuating real system, such a FFLO state is not realized as a phase-coherent SC phase. 


In Fig.5, the temperature dependences of the coefficients of the gradient terms at $H=H_{c2}(0)$, obtained in terms of the same set of the parameter values as in Fig.1, are shown. The fact that the coefficient $b_0$ is negative in the temperature range of Fig.5 implies that the FFLO modulation parallel to ${\bf H}$ may develop in the same temperature range within the mean field theory. On the other hand, it implies based on eq.(\ref{DOSmain}) that the $n=0$ LL term in $\Delta (\sigma_{\rm N})_{zz}|_0$ is positive so that the negative magnetoresistance behavior \cite{Kasa20} can never occur in any situation where the SC fluctuation is dominated by the $n=0$ LL one. In contrast, the corresponding coefficient $b_1$ is positive at such temperatures. Thus, the $n=1$ LL term in eq.(\ref{DOSmain}) is negative and becomes the origin of the remarkable negative magnetoresistance \cite{Kasa20}. 

\begin{figure}[b]
\begin{center}
{\includegraphics[scale=0.7]{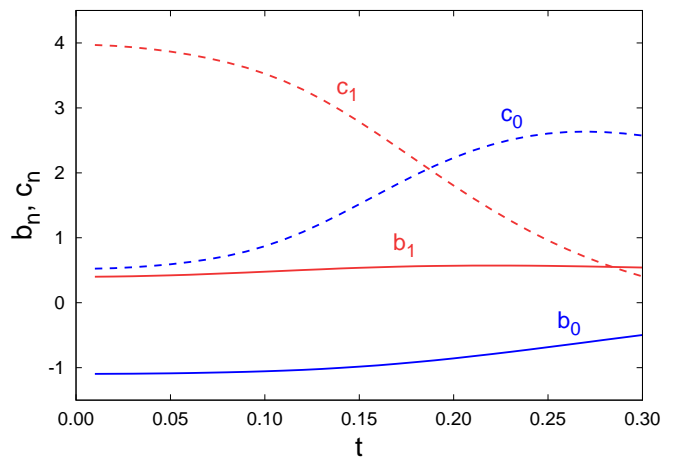}}
\caption{(Color online) Temperature dependences of the coefficients defined in eq.(\ref{grad}) at $H=H_{c2}(0)$, corresponding to $h=0.1806$ in Fig.1, which are obtained consistently with the $a_n(0)=0$ lines in Fig.1 and in terms of the parameter values listed below eq.(\ref{cons0}). 
} 
\label{Fig.5}
\end{center}
\end{figure}

Then, using eqs.(\ref{teikou}) and (\ref{r1}) and based on the procedures explained above, we have obtained resistivity curves using $T_{c0}/E_{\rm F} = 0.312$ which are presented in Fig.6. For simplicity, $H_{\rm VG}(T) = H_{c2}(0)(1 - t/2.8)$ was assumed in obtaining the resistivity curves numerically. 
Based on the understanding on Fig.1 mentioned above, let us focus hereafter on the fluctuation regime above the VG transition line in $T < 0.15 T_{c0}$ by neglecting the $n=0$ LL fluctuation. That is, just the $n=1$ contributions will be kept in eq.(\ref{AL}) and eq.(\ref{DOSmain}) 
as well as in eq.(\ref{VGcondfinal}). 
Then, the sign of the correction (\ref{DOSmain}) to the normal conductivity is negative and leads to negative magnetoresistance. It should be noted that this negative contribution to the conductivity is due to the $n=1$ LL fluctuation and is different in nature from that seen in 2D dirty limit \cite{GaL,JPSJ722930} in the quantum regime, because the latter is a feature of the $n=0$ LL fluctuation. 
\begin{figure}[t]
\begin{center}
{\includegraphics[scale=0.7]{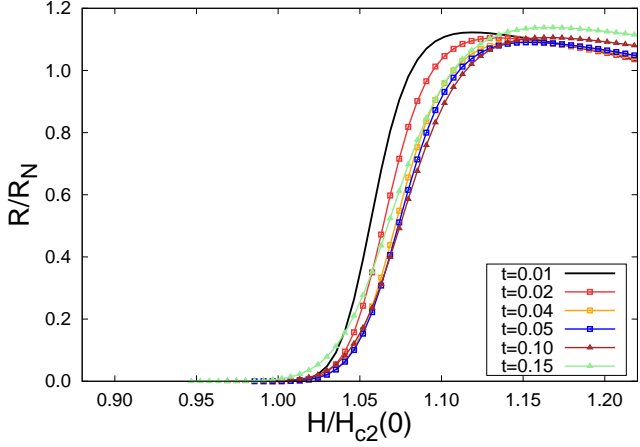}}
\caption{(Color online) Examples of the field dependences of the resulting curves of the normalized resistance, eq.(\ref{teikou}) with eq.(\ref{cons0}), at each value of the normalized temperature $t=T/T_{c0}$. We have used $T_{c0}/E_{\rm F}=0.312$ besides the parameter values used in obtaining Fig.5. 
} 
\label{Fig.6}
\end{center}
\end{figure}

The results in Fig.6 clearly show the two main features on the resistive transition in superconductors with strong PPB effects in the quantum regime we have emphasized in the preceding sections. First of all, the resistance {\it increases} on cooling over a wide field range except in the vicinity of $H_{c2}(0)$, reflecting $\sigma^{({\rm AL})}_{zz}$ and $\sigma^{({\rm VG})}_{zz}$ reducing upon cooling due to the quantum SC fluctuation. In Fig.6, the quantum fluctuation behavior of $\sigma^{({\rm AL})}_{zz}$ was enhanced by the large $\alpha_{\rm M}$ (i.e., strong PPB). We also note that the pinning strength $\Delta_{\rm P}=0.8$ used in Fig.6 is too large compared with the fluctuation strength. In fact, the resistance sufficiently decreases in higher fields than $H_{c2}$ there, and the feature in Fig.6 that the insulating behavior is remarkably seen is also a consequence of the large value of $\Delta_{\rm P}$ determining the magnitude of $\sigma^{({\rm VG})}_{zz}$. In any case, the details of the resistive curves in such low temperatures and high fields are quite sensitive to the functional form of $H_{c2}(T)$ at low $T$ which, in turn, is sensitive to the electronic Hamiltonian and the band structures. 

Next, the negative fluctuation conductivity correction (\ref{DOSmain}) is found to create negative magnetoresistance at lower temperatures like $T=0.01 T_{c0}$ and in the field range where the insulating behavior is seen. 

One might feel strange by looking at the high field behavior of the curves at higher temperatures, $t=0.1$ and $0.15$, which lie beyond the normal resistance even in higher fields than the fluctuation regime. This is a consequence of our neglect of the $n=0$ LL fluctuation. As Fig.5 shows, the negative sign of $b_0$ at lower temperatures implies that the $n=0$ LL contribution to $\Delta (\sigma_{\rm N})_{zz}|_0$ is positive, and hence, if the crossover from the $n=1$ LL regime to the $n=0$ LL regime is properly described, the resistance values at $t=0.1$ and $0.15$ would be suppressed below the normal resistance value. 

\subsection{$T_{c0}^{-1} \ll \tau_{\rm imp} \ll \tau_{\rm int}$}

In turn, we focus on the case in which the quasiparticle life time $\tau$ originates mainly from the impurity scattering event. In this case, regarding the correction term to the normal conductivity, eq.(\ref{cancelDOS}) should be used so that the total conductivity 
\begin{equation}
(\sigma_s)_{zz} = \sigma_{zz}^{({\rm AL})} + \sigma_{zz}^{({\rm VG})} + \Delta (\sigma_{\rm N})_{zz}  
\label{condcancel}
\end{equation}
needs to be used to obtain the resistivity curves. 

\begin{figure}[b]
\begin{center}
{\includegraphics[scale=0.7]{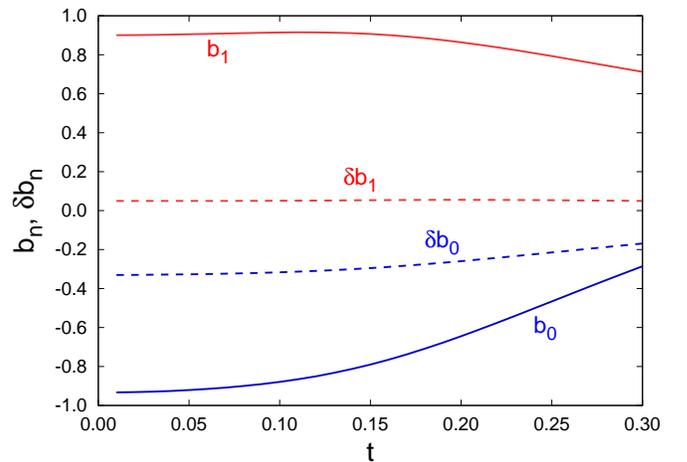}}
\caption{(Color online) Temperature dependences of the coefficients $b_n$ and $\delta b_n$ at $H=H_{c2}(0)$ obtained in terms of the parameter values ${\tilde J}=0.94$, $\gamma=1.05$, $\tau T_{c0}=50$, and $\alpha_{\rm M}=3.0$. 
} 
\label{Fig.7}
\end{center}
\end{figure}

To obtain the coefficients in the GL action in this case, we have used the following set of the parameter values, ${\tilde J}=0.94$, $\gamma=1.05$, $\tau T_{c0}=50$, and $\alpha_{\rm M}=3.0$. 
The PPB effect has been slightly reduced compared with that in the case A, as can be seen from the smaller values of $\tau T_{c0}$ and $\alpha_{\rm M}$. Nevertheless, the $n=1$ LL modes dominate over the $n=0$ LL ones, like in Fig.1, at least in $t < 0.1$. The resulting temperature dependences of the coefficients appearing in eqs.(\ref{DOSmain}) and (\ref{cancelDOS}) are shown in Fig.7. Since the coefficient $b_1$ remains positive in the low temperature range of our interest, eq.(\ref{r1}) with $n=1$ can be used for the fluctuation renormalization. 

To examine the resistivity curves in a current parallel to the field \cite{Kasa20} in terms of eqs.(\ref{teikou}) and (\ref{condcancel}), we have used the parameter values $\Delta_{\rm P}=0.02$ and $T_{c0}/E_{\rm F}=0.712$ together with the relation $H_{\rm VG}(T) = H_{c2}(0) ( 1 - t/1.8)$ and the parameter values used for Fig.7. The small $\Delta_{\rm P}$-value indicates that the weight of $\sigma_{zz}^{({\rm VG})}$ in eq.(\ref{condcancel}) and thus, the quantum fluctuation contribution to $(\sigma_s)_{zz}$ is reduced in this case. Further, to make negative magnetoresistance arising from the positive $\delta b_1$ (see Fig.7) visible, we have assumed here an unusually large $T_{c0}/E_{\rm F}$-value. 

In the case A, we have noted that the $n=0$ LL fluctuation does not lead to the negative magnetoresistance of the type seen in FeSe below 1 (K) \cite{Kasa20} due to the negative sign of $b_0$. Similarly, even in this case, the fact that $\delta b_0$ remains negative as well as $b_0$ in the temperature range of our interest clarifies based on the expression (\ref{cancelDOS}) that the $n=0$ LL fluctuation cannot become the origin of negative magnetoresistance at the low temperatures. This conclusion is not changed even in the case with a much lower value of the Maki parameter $\alpha_{\rm M}$ where $\delta b_0$ at lowest temperatures remains positive, because $\delta b_0$ generally increases upon warming, and consequently, it is difficult to find, as in the experimental data \cite{Kasa21}, the situation with positive magnetoresistance at higher temperatures. 

The resulting resistivity curves obtained in terms of the above-mentioned parameter values in the present case B are shown in Fig.8. Taking account of the fact that the temperature region in which the $n=1$ LL vortex lattice is stable is narrower in this case, just the resistivity curves at the lowest three temperatures are presented. The insulating behavior and the negative magnetoresistance are clearly visible in $H_{c2}(0) < H < 1.04 H_{c2}(0)$. In addition, the negative sign of $\delta b_0$ in Fig.7 indicates that some disappearance of the negative magnetoresistance at intermediate temperatures \cite{Kasa20} would be realized by including the $n=0$ LL fluctuations in our numerical calculations. The inclusion of the $n=0$ LL modes should be performed consistently with our future study on the crossover between the two LL fluctuations and the transition between the HFSC phase, i.e., the $n=1$ LL vortex lattice, and the ordinary $n=0$ LL triangular vortex lattice. 

\begin{figure}[t]
\begin{center}
{\includegraphics[scale=0.7]{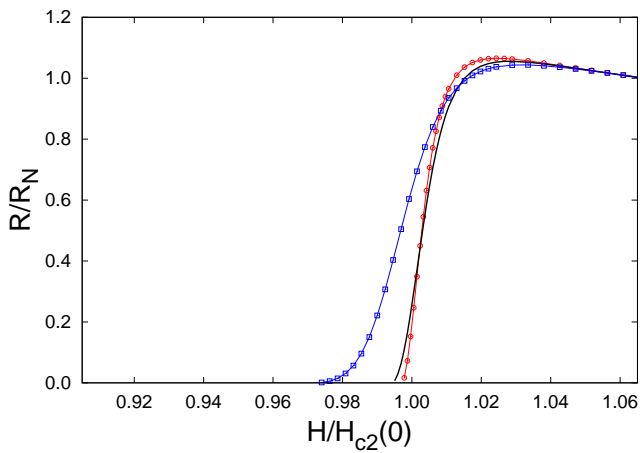}}
\caption{(Color online) Examples of the field dependences of the resulting curves of the normalized resistance, eq.(\ref{teikou}) with eq.(\ref{condcancel}), at the normalized temperatures $t=T/T_{c0}=0.05$ (blue square), $0.01$ (black solid curve), and $0.005$ (red circle). We have used $T_{c0}/E_{\rm F}=0.712$ besides the parameter values listed in the text and the caption of Fig.7. 
} 
\label{Fig.8}
\end{center}
\end{figure}

\section{Summary and Discussion}

In the present work, the fluctuation conductivity in a superconductor with moderately strong PPB has been examined by focusing on the low temperature and high field region, and the obtained results have been discussed in relation to the strange resistivity behavior seen in the fluctuation regime of FeSe in a field parallel to the basal plane. It has been argued that the sign of the sum of the DOS and MT contributions to the fluctuation conductivity is the origin of the remarkable negative magnetoresistance in FeSe at low temperatures in a current parallel to the field \cite{Kasa20}. Based on this observation, the novel high field SC phase of FeSe in a field parallel to the basal plane has been argued to be a pinned vortex lattice described by the SC order parameter in the $n=1$ Landau level (LL) and not the familiar FFLO vortex lattice with a modulation parallel to the field which is described by the order parameter in the $n=0$ LL. Further, an enhancement of the quantum fluctuation due to PPB has been pointed out and identified as the origin of the insulating behavior of the resistivity, which is another feature seen in FeSe in the fluctuation regime \cite{Kasa20}. 

Realization of a higher LL vortex state in FeSe is, broadly speaking, due to relatively strong PPB in the parallel field configuration. In the perpendicular field configuration (${\bf H} \parallel c$), on the other hand, a different HFSC phase was argued to be present in FeSe \cite{Kasa21} and has been recently identified as a $n=0$ LL vortex lattice with a spatial modulation parallel to the field \cite{NI}. In fact, the $\alpha_{\rm M}$-value in ${\bf H} \parallel c$ case has been estimated to be below 1.0 which is much smaller than those used in the present work.  

As already mentioned, the $d$-wave pairing has been assumed in obtaining Fig.6 and 8. In contrast, in the conventional $s$-wave pairing, the negative magnetoresistance in the fluctuation regime does not seem to occur. In addition, we note that an unexpectedly large $T_{c0}/E_{\rm F}$-value seems to be necessary to obtain such a sizable negative magnetoresistance theoretically. It suggests that the superconductivity in FeSe cannot be described fully within the BCS framework \cite{MatsudaHanaguriShibauchi}, and, in addition, that the present simplest model based on a single-band weak-coupling BCS Hamiltonian is insufficient for description. The corresponding analysis based on a two-band model \cite{Mizukami,Adamaster} appropriate for describing FeSe is needed and will be left for our future work.

\section{Acknowledgement}
One of the authors (R.I.) thanks S. Kasahara and Y. Matsuda for discussions on their experimental data. The present work was supported by KAKENHI (No.21K03468) from Japan Society for the Promotion of Science.

\section{Appendix A} 

Derivation of eq.(\ref{appen}) will be presented in this Appendix. Since we have shown in eq.(\ref{isop}) how to rewrite the expressions in the isotropic case into the corresponding ones in the anisotropic case under the condition (\ref{gammaMaki}), we focus below on the expressions in the isotropic case. 

In obtaining eq.(\ref{appen}), the dependences on $q$ and the frequencies are trivially understood, and the order parameter in the $n$-th LL will be expressed, in place of eq.(\ref{LLeigen}), as $\Delta_n(x,y)$ under the gauge ${\bf A}= - H y {\hat x}$. Then, the annihilation operator ${\hat a}$ satisfying the commutation relation ${\hat a} {\hat a}^\dagger - {\hat a}^\dagger {\hat a}=1$ will be chosen as 
\begin{equation}
{\hat a}=\frac{r_H}{\sqrt{2}}(\Pi_x - {\rm i} \Pi_y).
\end{equation}
Here, let us introduce the coherent state for the LLs 
\begin{equation}
\Delta^{(\sigma)}(x,y) \equiv e^{-|\sigma|^2/2} e^{\sigma {\hat a}^\dagger} e^{-\sigma {\hat a}} \Delta_0(x,y). 
\label{coh1}
\end{equation}
Using the formula 
\begin{equation}
e^{{\hat A}+{\hat B}} = e^{\hat A} e^{\hat B} e^{-C/2}
\label{BCH}
\end{equation}
with a constant $C= {\hat A} {\hat B} - {\hat B} {\hat A}$, eq.(\ref{coh1}) is rewritten as 
\begin{equation}
\Delta^{(\sigma)}(x,y) = e^{(\sigma^2 -|\sigma|^2)/2} \Delta_0(x, y+\sqrt{2}\sigma), 
\label{ident1}
\end{equation}
where $\sigma$ is a complex constant. On the other hand, using the formula (\ref{BCH}) twice, one can verify the relation 
\begin{equation}
\exp(\nu {\hat a}^\dagger - \nu^* {\hat a}) \Delta^{({\overline \sigma})}(x,y) = \exp\biggl(\frac{{\overline \sigma}}{2} (\nu - \nu^*) \biggr) \Delta^{({\overline \sigma} + \nu)}(x,y), 
\label{ident2}
\end{equation}
where ${\overline \sigma}$ is a real constant, and $\nu$ is a complex constant. Noting that ${\rm i}\rho({\bf v}\cdot{\bf \Pi})/T_{c0} = \rho(\mu {\hat a}^\dagger - \mu^* {\hat a})$ where $\mu$ is given by eq.(\ref{mu}) and using the relations (\ref{ident1}) and (\ref{ident2}), we have the following relation 
\begin{widetext}
\begin{equation}
\exp\biggl({\rm i}\rho\frac{{\bf v}\cdot{\bf \Pi}}{T_{c0}} \biggr) 
\Delta^{({\overline \sigma})}(x,y) 
= \exp\biggl(\rho(\mu - \mu^*){\overline \sigma} + \frac{\rho^2}{2} (\mu^2 - |\mu|^2) \biggr) \Delta_0(x, y+\sqrt{2}({\overline \sigma}+\mu \rho)).
\end{equation}
\end{widetext}
Then, as the O(${\overline \sigma}^n$) ($n=0$, $1$) contributions of the above expression, we obtain 
\begin{eqnarray}
\exp \biggl({\rm i}\rho \frac{{\bf v}\cdot{\bf \Pi}}{T_{c0}} \biggr) \Delta_0(x,y) &=& \exp \biggl(\frac{{\rho^2}}{2} (\mu^2 - |\mu|^2) \biggr) \nonumber \\ 
&\times& \Delta_0(x, y + \sqrt{2} \mu \rho), \nonumber \\
\exp \biggl({\rm i}\rho \frac{{\bf v}\cdot{\bf \Pi}}{T_{c0}} \biggr) \Delta_1(x,y) &=& \exp \biggl(\frac{{\rho^2}}{2} (\mu^2 - |\mu|^2) \biggr) \biggl(\rho(\mu-\mu^*) \nonumber \\
&+& \frac{\partial}{\partial(\mu \rho)} \biggr) \Delta_0(x, y + \sqrt{2} \mu \rho). 
\end{eqnarray}
Finally, by performing the spatial integrals for the above two expressions multiplied by $\Delta_n^*(x,y)$ from the left, one obtains eq.(\ref{appen}). 
 
\begin{figure}[t]
\begin{center}
{\includegraphics[scale=1.9]{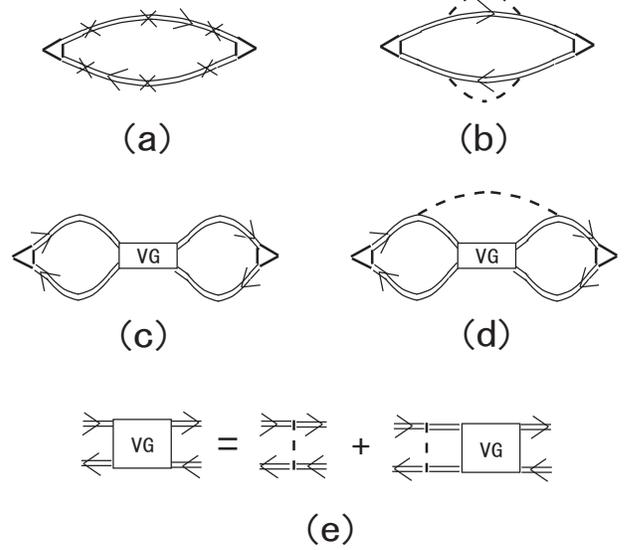}}
\caption{(Color online) (a): Diagram expressing $\sigma^{({\rm AL})}_{zz}+\sigma^{({\rm VG})}_{zz}$ prior to the pinning potential average. A cross symbol denotes a pinning potential. (b): Diagram correcting $\sigma^{({\rm AL})}_{zz}$ arising from (a) after the pinning potential average. (c) and (d): Diagrams associated with $\sigma^{({\rm VG})}_{zz}$ arising from (a) after the pinning potential average. (e): Diagram on the VG fluctuation propagator expressed within the ladder approximation. The thick dashed lines in (b), (d), and (e) denote the pinning potential line carrying the strength $\Delta_{\rm P}$. 
} 
\label{Fig.9}
\end{center}
\end{figure}

\section{Appendix B} 

In this Appendix, the formulation \cite{RI97} leading to identifying Fig.3 as the diagram describing the VG scaling behavior is briefly explained. 
Based on the GL action (\ref{GLaction}) with eq.(\ref{pinning}), the AL fluctuation conductivity defined prior to the pinning potential average can be regarded as being of the form of Fig.9 (a), where the cross symbols denote the pinning potential $U({\bf r})$. After the pinning potential average, the figure (a) results in the figures (b), (c), (d), and Fig.3. Here, the figure (b) can be regarded as one part of the diagram expressing $\sigma^{({\rm AL})}_{zz}$, while other three diagrams can lead to a different temperature dependence in the conductivity through the VG fluctuation propagator sketched in the figure (e). Clearly, the figure (c) vanishes. However, the figure (d) is found to lead to a divergent contribution, proportional to $ T \Delta_{\rm P}^2 {\overline \xi}_{\rm VG}^{z-3}$, to the conductivity at finite temperatures. On the other hand, as is shown in sec.II, Fig.3 leads to a more divergent contribution proportional to $T \Delta_{\rm P}^3 {\overline \xi}_{\rm VG}^{z-1}$. Although the former may become dominant for a weaker pinning strength far from the VG criticality, we have chosen only the latter, which is of a higher order in $\Delta_{\rm P}$ but is more divergent on approaching the VG criticality, in performing the numerical calculations leading to Figs.6 and 8.

\end{document}